\documentclass[12pt]{article}
\usepackage{latexsym}
\usepackage{epsfig,amssymb,euscript}
\usepackage{amsmath}
\usepackage{bbm}

\def\be{\begin{equation}}
\def\ee{\end{equation}}
\def\bea{\begin{eqnarray}}
\def\eea{\end{eqnarray}}

\input epsf

\begin{document}

\begin{titlepage}
\font\cmss=cmss10 \font\cmsss=cmss10 at 7pt
\hfill SISSA-34/2006/EP
\\
\vspace{25pt}

\begin{center}
{\LARGE \bf On Stable Non-Supersymmetric Vacua}
\vskip 20pt
{\LARGE \bf at the Bottom of Cascading Theories}

\end{center}

\vspace{15pt}

\begin{center}
{\large  Riccardo Argurio$^1$, Matteo Bertolini$^2$, }
\vspace{7pt}

{\large Cyril Closset$^1$ and Stefano Cremonesi$^2$}\\

\vspace{25pt}
{\sl $^1$Physique Th\'eorique et Math\'ematique \\
and International Solvay Institutes \\
Universit\'e Libre de Bruxelles \\
CP 231, 1050 Bruxelles, Belgium}

\vspace{15pt}
{\sl $^2$SISSA/ISAS and INFN - Sezione di Trieste\\
Via Beirut 2; I-34014 Trieste, Italy}

\end{center}

\vspace{20pt}

\begin{center}
\textbf{Abstract}
\end{center}
We consider a wide class of cascading gauge theories which usually lead to runaway behaviour 
in the IR, and discuss possible deformations of the superpotential at the bottom of the cascade 
which stabilize the runaway direction and provide stable non-supersymmetric vacua. The models we find 
may allow for a weakly coupled supergravity analysis of dynamical supersymmetric breaking in the context 
of the gauge/string correspondence.

\end{titlepage}

\renewcommand{\thefootnote}{\arabic{footnote}}
\setcounter{footnote}{0} \setcounter{page}{1}


\tableofcontents

\section{Introduction and Summary}

A new approach to study the strongly coupled dynamics of gauge theories
has been provided by their holographic correspondence with string theory
on specific backgrounds (for reviews, see for instance 
\cite{Aharony:2002up,Bertolini:2003iv,Bigazzi:2003ui}, and \cite{Imeroni:2003jk,Paredes:2004xw} 
for PhD thesis). In particular, with this approach one can in principle have access to 
properties of the theory, such as the full low-energy spectrum, which
are beyond one's reach even when dealing with supersymmetric gauge theories.

Focusing on ${\cal N}=1$ supersymmetric (SUSY) gauge theories, what on the other hand is
within reach using purely gauge theoretic methods, is the information about
the holomorphic quantities of the theory, in particular the number of isolated 
exact vacua, or the quantum moduli space of vacua. 
Paradoxically, it is possible by the same methods to argue
that the theory has no supersymmetric vacua because of the non-perturbative
dynamics \cite{ADS83vc,ads}. There are
then two possibilities. Either one approaches a SUSY vacuum as some VEVs
become infinite, which is referred to as having a runaway behaviour (the
theory does not have a proper vacuum), or there
is a stable non-supersymmetric vacuum, which we will refer to here
as dynamical supersymmetry breaking (DSB). The latter option is obviously
the favoured one when there is no moduli space at the classical level.

However, besides arguing that SUSY is broken at hierarchically small scales 
\cite{Witten:1981nf}, we do not have the tools to really analyze quantitatively the physics
around the vacuum.
It would thus be extremely interesting to have at hand an example
of gauge theory which displays DSB and which has a string/gravity dual.
Keeping aside the usual problems related to the decoupling of the UV 
completion of the theory in the gravity dual, we would in principle be able 
to compute the spectrum of low-energy fields, their interactions, and so on.

A recent progress in the AdS/CFT correspondence was the discovery of new infinite classes 
of Sasaki-Einstein (SE) manifolds $X^5$ which, through considering type IIB string theory 
on $AdS_5 \times X^5$, correspond to some ${\cal N}=1$ four-dimensional superconformal gauge 
theories, engineered placing a bunch of $N$ D3 branes at the tip of the CY cone over $X^5$. These 
new manifolds were dubbed $Y^{p,q}$ \cite{gmswSE}, $L^{p,q,r}$ \cite{Cvetic:2005ft,Martelli:2005wy} 
and $X^{p,q}$ \cite{Hanany:2005hq} (of the latter the explicit metrics are not known). The dual 
quiver gauge theories were constructed and many checks of the correspondence were carried out 
\cite{MS,MS2,bbc,Benvenuti:2005ja,Butti:2005sw,Franco:2005sm}.

A subsequent progress, which is more important to our purposes here, came adding fractional branes 
to these systems. Fractional branes are usually meant to break conformal 
invariance and move towards more realistic and dynamically interesting theories. In particular, they trigger a 
renormalization group flow which takes
the form of a cascade of Seiberg dualities 
reducing the ranks of the gauge groups as we flow towards the IR,
as first discussed for the conifold case \cite{KT,KS,Strassler:2005qs}. 
The cascade goes on until for some gauge group(s) the number of flavours 
equals the number of colours. At this point the quantum moduli space of the corresponding gauge group(s) gets 
modified. The theory confines along the baryonic branch and the full theory is reduced by one (or more) 
gauge group factor. The 
theory stops cascading as the bottom of the cascade has been reached. If one looks at the structure of the gauge 
theory within this energy range, it is as if regular branes have completely disappeared and only fractional branes 
have survived. The IR dynamics of the theory is then determined solely by the fractional branes. For a detailed review 
on the physics of the cascade we refer the reader to the beautiful paper \cite{Strassler:2005qs}.

Contrary to the case of the conifold \cite{KS}, where the cascade
ends up with confining SUSY vacua, 
fractional branes on these newly found SE manifolds 
end up at the bottom of the cascade with the generation of an 
Affleck-Dine-Seiberg (ADS) superpotential, 
and hence generically break supersymmetry. 
It has by now become clear that there exist three classes
of fractional branes in string theory, according to the
different IR dynamics they induce in the dual gauge theory \cite{FHSU}:
{\it i)} ${\cal N}=2$ branes, which correspond to branes wrapped on cycles 
that are not located at a point-singularity but rather on a curve singularity and hence 
have a moduli space and generate ${\cal N}=2$ supersymmetric dynamics;
{\it ii)} {\it deformation} branes, which trigger confinement and lead 
to supersymmetric vacua; {\it iii)} {\it supersymmetry breaking} (SB) branes, which generate an ADS superpotential 
and might lead to stable non-supersymmetric vacua or runaway behavior in the dual 
gauge theory.

In all the cases with SB branes that have been studied a runaway direction develops 
\cite{BHOP,FHSU,bbc2,IS,Brini:2006ej,Butti:2006hc}. 
Therefore, the possibility of getting fully stable non-supersymmetric vacua, in the context 
of the gauge/string correspondence, remains an open challenge.
\footnote{The possibility of meta-stable long-lived vacua was pointed out recently for SQCD with 
massive \cite{Intriligator:2006dd} and massless \cite{Franco:2006es} flavors. These models find a nice  
realization into flavoured versions of AdS/CFT \cite{Franco:2006es}. See also
\cite{ooguri} for more work on meta-stable DSB.}

In this paper, we address this issue and show how in some cases
simple deformations of the tree level superpotential can
change the IR dynamics in such a way that stable
non-supersymmetric vacua do arise. This is not generic and holds
only in very specific cases. This pairs with the fact that
DSB models with stable vacua arise only for very
specific choices of gauge groups and matter content, while
generically one expects either supersymmetric vacua or runaway
behaviour. It would be nice to understand this property directly
from string theory. We perform a pure field theory
analysis, and do not attempt to answer this question here, though some
comments are given in the discussion section.

Our analysis suggests that a generic, almost model-independent pattern for 
DSB emerges. We consider, as explicit examples,  
the gauge theories on fractional branes placed at the tip of the complex cone over toric 
del Pezzo surfaces, $dP_k$. In fact, these are related to the SE manifolds recently found. 
In particular, the complex cone over the first del Pezzo is the real cone over 
$Y^{2,1}$ \cite{MS}, and the complex cone over the second del Pezzo is the real cone 
over $X^{2,1}$ \cite{Hanany:2005hq}. Hence the corresponding dual gauge theories are also the same  
(for earlier works on regular branes at del Pezzo conical singularities, see 
\cite{Wijnholt:2002qz,Herzog:2003zc}). The metrics for the CY cone over the del Pezzo surfaces are not known, 
but for $dP_1$. However, the toric diagrams are known. This is enough to derive the dual gauge theory 
\cite{Franco:2005rj,Hanany:2005ss,Feng:2005gw}. Support for a cascading dynamics for the gauge theories 
of fractional branes at del Pezzo's conical singularities was given by supergravity analysis in 
\cite{Franco:2004jz,HEK}. For a gauge theory analysis confirming these expectations we refer to 
\cite{FHU,BHOP,FHSU,bbc2}. In what follows, we will always assume that a cascade takes place and analyze 
the gauge theory directly at the last step, i.e. in the deep IR. 

The paper is organized as follows.

In section 2 we briefly review known results about the $dP_1$ case and partially extend them. We show, 
in particular, that stable non-supersymmetric vacua arise only when perturbing the superconformal 
fixed point by a single fractional brane while for a generic number $M$ of fractional branes 
no deformations are possible which lift
all classical flat directions and cure the runaway behaviour. We 
also argue that it is not possible to obtain a stable
non-supersymmetric vacuum in the full class of $Y^{p,q}$ manifolds,
independently on the number $M$ of fractional branes. 

In section 3 we consider the case of the second del Pezzo surface. There are two 
kinds of fractional branes here: deformation branes and SB branes. Similarly to 
the previous case, the latter induce a runaway behaviour. The generic situation is having $P$
deformation branes and $K$ SB branes. Similarly to the $dP_1$ case, the addition of
one (and only one) SB brane, $K=1$, allows for a deformation of the
superpotential which stabilizes the runaway direction and provides a
stable non-supersymmetric vacuum at finite field VEVs. There is a crucial 
difference with respect to the $dP_1$ case, though. The DSB model we find here allows for a large number of 
deformation branes. This makes it possible to perform a probe analysis of a single SB brane in the weakly 
coupled and well-behaved background generated by the deformation branes. We do not pursue 
this analysis here, nevertheless in principle a dual supergravity description 
is possible in this case.

Section 4 contains the analysis for $dP_3$. The story repeats:
generically, it is not possible to find deformations of the
superpotential that stabilize the runaway direction and get a
non-supersymmetric vacuum. However, for a specific choice of
fractional branes (and only for such a choice) we recover a model very similar to the 
one working for $dP_2$. Once again, a large number of 
deformation branes may support a smooth and weakly coupled background where 
to perform a probe analysis.

Section 5 contains a discussion for other del Pezzo toric varieties,
the so-called pseudo del Pezzo's \cite{Feng:2002fv}. We focus for definiteness 
on $PdP_4$ and show that similar phenomena to those occurring for $dP_2$ and $dP_3$ hold.

We conclude in section 6 with a discussion. Our analysis indicates that a generic 
pattern for DSB does emerge. More precisely, the actual models of DSB always turn out to be 
a minimally modified version of the $SU(N)-SU(2)$ model \cite{pst}. 
Unfortunately, we cannot yet provide a direct string theory explanation
of this result, but the possibility of performing a probe analysis 
of the DSB phenomenon makes it worth trying to find KS-like solutions 
corresponding to these geometries.

\section{Stable Vacua at the Bottom of the $\mathbf{dP_1}$ Cascade}

Let us start by analyzing the simplest case, namely the dynamics of a set of 
$M$ fractional D3 branes on the complex cone over
the first del Pezzo surface, $dP_1$. 
In this case there is only one type of fractional brane
which is of the SB type \cite{BHOP,FHSU,bbc2}. The IR dynamics
of this model has been analyzed in detail in the
literature, showing that it displays a runaway behaviour along a
baryonic flat direction. In \cite{IS}, it was also noted that for a 
single fractional brane ($M=1$) the runaway direction can in fact be converted 
into a stable non-supersymmetric vacuum via some modification of the tree level
superpotential. The theory at the bottom of the cascade gets essentially
reduced to the 3-2 model of \cite{ads}.
Here, we briefly repeat the
analysis to set up the framework for the more general cases discussed later.

\begin{figure}[ht]
\begin{center}
{\includegraphics{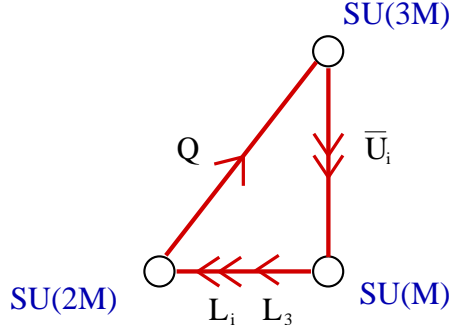}}
\caption{\small Quiver of the $dP_1$ theory for $M$ fractional branes.}
\label{ldp1g}
\end{center}
\end{figure}

The theory at the bottom of the cascade is depicted in the quiver in
Figure \ref{ldp1g}.
There is in addition a tree-level superpotential 
\be
\label{wtree}
W_{\mathrm{tree}} = h~ Q\,\bar{U_i} L_j\, \epsilon^{ij} ~~~~~i=1,2~,
\ee
while the R-charges of the chiral superfields are reported in the
table below (the $\overline{U}_i$'s and $L_i$'s are both in a doublet of
the global $SU(2)$ and hence must have the same R-charge).
\begin{table} [ht]
\label{Rcdp1}
\begin{center}
\begin{tabular}{c|c|c|c|c|}
 & $Q$ & $\overline{U}_i$ & $L_i$ & $L_3$ \\
\hline
$U(1)_R$ & -1  & 0 & 3 & -1
\end{tabular}
\end{center}
\end{table}

This theory has a number of classical flat directions. It is easy to see
that the F-flatness conditions derived from the superpotential (\ref{wtree}) set all the 
invariants involving the fields $Q$ and $\overline{U}_i$ to zero.
However there are a number of ``baryonic'' invariants, schematically written
as
\be
{\cal B}_{i_1\dots i_k} = 
\epsilon_{2M}\epsilon_M \epsilon_M (L_i)^{2M-k}(L_3)^k~~, \qquad
\qquad k=0,\dots, M~,
\label{baryons}
\ee
where $\epsilon_{2M}$ and $\epsilon_M$ are the Levi-Civita tensors of $SU(2M)$ and
$SU(M)$, respectively. One can actually see that if we consider the $L_a=(L_i,L_3)$ to be in a
triplet $\mathbf{3}$ of a global (accidental) $SU(3)$, 
then all the invariants above
fall into one $SU(3)$ representation which is the $M$th symmetric product
of the $\bf \bar 3$, of dimension $\frac{1}{2}(M+1)(M+2)$.
This representation splits into a sum of $SU(2)$ symmetric representations
with $k$ indices, for $k=0,\dots, M$, corresponding to the ${\cal B}_{[k]}$ 
above.

As explained in \cite{IS}, at a generic point of the moduli space spanned
by the ${\cal B}_{[k]}$'s, the superfields $Q$ and $\overline{U}_i$ get a mass, 
and the determinant of the mass matrix is proportional to ${\cal B}_{[0]}$
(the baryon without $L_3$'s, which is the global $SU(2)$ singlet). Hence, 
via the matching of scales and the generation of a non-perturbative 
superpotential for the confining $SU(3M)$ gauge group, we have a runaway
behaviour along the direction parameterized by ${\cal B}_{[0]}$.\footnote{
The K\"ahler potential, which is approximated by the classical one for 
large enough VEVs, can be shown to be consistent with this conclusion.}

The above reasoning is true for any number $M$ of fractional branes.
We can now ask whether or not it is possible to lift the classical flat 
directions by deforming the tree level superpotential with the addition
of baryonic couplings. Indeed, lifting the classical flat directions
is a first step towards finding a model which realizes DSB.

Another crucial ingredient for obtaining DSB is the following.
A well known criterion
\cite{ADS83vc,ads} for the possibility of having a stable
non-supersymmetric vacuum in a supersymmetric gauge theory is the
presence of a non-anomalous R-symmetry which is
spontaneously broken by quantum effects. If a theory realizes this
condition and does not have classical flat directions, one expects
the full quantum theory to develop a stable non-supersymmetric
vacuum.\footnote{Indeed, the well-known argument is that 
SUSY must also be broken since there
is no room for a flat non-compact direction parameterized by a
putative scalar superpartner of the Goldstone boson associated to the
breaking of the R-symmetry.}

The theory at the bottom of the $dP_1$ cascade,
figure \ref{ldp1g} and superpotential (\ref{wtree}), 
has indeed a non anomalous
R-symmetry, independently of the value of $M$. 
When one aims to suitably deform the theory in order to lift the classical
flat directions, it is crucial to check that the extra contributions
to the tree level superpotential respect this symmetry. In other words, 
the operators one wants to add to $W_\mathrm{tree}$ must have R-charge 2.

\begin{figure}[ht]
\begin{center}
{\includegraphics{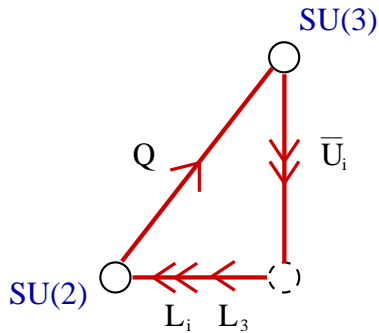}}
\caption{\small Quiver of the $dP_1$ theory for one fractional brane. 
The dashed node represents an ``SU(1)'' factor, which plays no r\^ole in the IR.}
\label{ldp1}
\end{center}
\end{figure}

The R-charge of the baryonic operators ${\cal B}_{[k]}$ is simply
\be
R({\cal B}_{[k]})=6M-4k\geq 2M.
\ee
The smallest R-charge is provided by the baryon with most $L_3$'s, in the
largest $SU(2)$ representation, i.e. ${\cal B}_{[M]}$. This
R-charge can be equal to 2 only if $M=1$. Hence, only in the presence
of a single fractional brane we expect to have a deformation of 
$W_\mathrm{tree}$ that leads to DSB. In \cite{IS} it was shown
that this is really the case. Indeed, the superpotential
\be
\label{wdef}
W_{\mathrm{def}} = W_{\mathrm{tree}} + \alpha  {\cal B}_1= h~  Q\,\bar{U_i} L_j\, \epsilon^{ij} + \alpha L_1 L_3
\ee
preserves the R-symmetry and effectively reduces the theory to the 3-2 model
of \cite{ads}, which displays DSB. The theory with $M=1$ is reported in figure \ref{ldp1}

It can be checked that for $M>1$ it is not possible to have DSB.
For $M=2$, a straightforward but tedious computation reveals
that, if one succeeds in lifting all the classical flat directions, the effect
of the quantum dynamics is just to displace the SUSY vacua at finite distance. 
For $M>2$, one can easily see that at the classical level it is
impossible to write simple couplings (i.e. linear in the baryonic operators)
which lift all the flat directions. Hence at the quantum level we will 
eventually get either runaway behaviour or moduli spaces of SUSY vacua.

For future reference let us finally notice that in what discussed above, a crucial ingredient for 
being able to write the extra term in the
superpotential (\ref{wdef}), was to have bi-fundamental
fields between an $SU(2)$ factor and an ``$SU(1)$'' factor.\footnote{
In this paper we consistently neglect the presence of possible $U(1)$ gauge
factors on every node. Besides the diagonal one which decouples completely,
we will assume that all the others get a mass through a GS mechanism, 
whether or not there are anomaly free combinations.}
This basically
allows one to ``turn around'' the arrow in forming a quadratic gauge invariant.

To conclude, the main lesson of the $dP_1$
case is that, while for any number $M$ of fractional branes the IR dynamics
is runaway, only for $M=1$ it is possible to cure this behaviour 
and achieve DSB. This is done 
by adding a suitable term to the tree level superpotential.

We know from general considerations in the gauge/gravity 
correspondence that having only one fractional brane is the opposite
regime with respect to the one for which the supergravity approximation
is really under control. In the present example, we cannot even argue
that the single fractional brane could be treated in the probe 
approximation, since the same fractional 
brane is at the same time also responsible for the 
confining dynamics at the $SU(3)$ node,  
which should correspond in some way to a deformation of the dual
geometry removing the singularity at the apex of the cone. 
This dynamics would clearly be missed in a probe approximation.
We will be able to circumvent this difficulty by considering the theory
based on $dP_2$ (and higher del Pezzo's).

\subsection{Comments on the $Y^{p,q}$ Family}

As already reminded, the complex CY cone over $dP_1$ is the same as the real CY cone over 
$Y^{2,1}$ \cite{MS}, this being the simplest instance of the infinite $Y^{p,q}$ family. 
The corresponding gauge theory  duals have been widely studied in the last two years and all results, so
far, indicate that the basic physical properties are quite the same, both in the conformal and in
the non-conformal cases, regardless the specific value of $p$ and $q$. Hence, one could ask if a story 
similar to the one just discussed holds for any $Y^{p,q}$. In what follows, we argue this is not the case.

The physics of the gauge theory dual to the $Y^{p,q}$ manifolds with fractional branes is believed to be a 
cascading theory, as it is the case for $Y^{2,1}$. There is a single type of fractional branes leading to a 
unique non-conformal quiver \cite{MS2}. Similarly to $Y^{2,1}$ 
the gauge theory at the bottom of the cascade has a classically flat baryonic runaway direction 
\cite{Brini:2006ej}\footnote{The details of the cascade are fully understood only for $q=1,p-1$. However, 
the authors of \cite{Brini:2006ej} have given convincing arguments for the runaway nature of the whole family.}.

The more $p$ and $q$ are increased, the more the analysis gets involved. However, explicit computations for 
$Y^{3,1}$ and $Y^{3,2}$ show that, independently on the number of fractional branes, the runaway direction 
cannot be lifted. Analyzing the structure of the corresponding quivers one sees that also for $M=1$ there are not 
enough matter fields charged under an $SU(2)$ and an ``$SU(1)$'' factor, this being a crucial ingredient 
to make a DSB deformation possible. Furthermore, there are several mesonic trace operators absent from the 
superpotential which are difficult to lift. 
This is a common feature for the whole $Y^{p,q}$ family. Hence, we expect that for $p$ and $q$ not being equal 
to 2 and 1, respectively, it is not possible to deform the fractional brane gauge theory and get stable 
non-supersymmetric vacua \cite{ABC}.

\section{Stable Vacua at the Bottom of the $\mathbf{dP_2}$ Cascade}

As we have seen in the previous section, in the $dP_1$ case we have found
only one instance where DSB could be provoked. In a sense, it would
be nicer to have a whole class of such models, where by tuning one
parameter we can place ourselves in a regime more suitable to study
the gravitational counterpart. For this reason, we turn to study fractional 
branes at the tip of the complex cone over the second del Pezzo surface, 
$dP_2$. This system was widely studied recently, see \cite{FHSU,Pinansky}.

\begin{figure}[ht]
\begin{center}
{\includegraphics{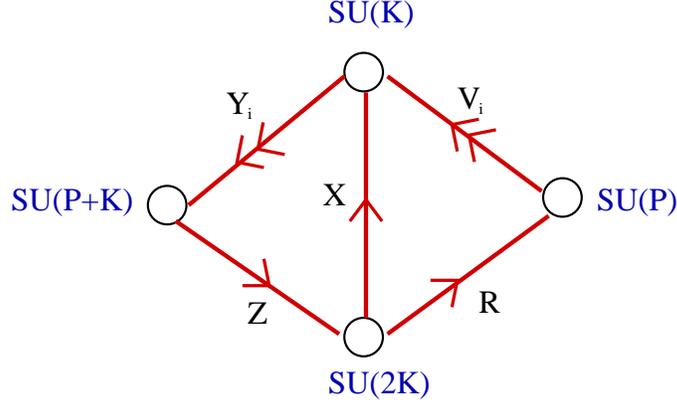}}
\caption{\small Quiver of the $dP_2$ theory for generic $P$ and $K$.}
\label{ldp2g}
\end{center}
\end{figure}

In $dP_2$, there are two kinds of fractional branes, one of deformation type
and one of supersymmetry breaking type. In Figure \ref{ldp2g} we show the quiver
diagram one gets at the end of the cascade with $P$ deformation
branes and $K$ SB branes. The cascade in this case is more subtle
than in the $dP_1$ case, for instance it is self-similar only after more
than one step of Seiberg dualities, and the number of steps after which
the pattern repeats itself depends on the relative size of $P$ and $K$.
However, one can check that in all the cases of interest one indeed finds
the quiver of Figure \ref{ldp2g} at the bottom of the cascade, with
the relevant values of $P$ and $K$.

The tree level superpotential inherited from the conformal
quiver is
\be
W_\mathrm{tree}= - Y_1ZX+ Y_2ZRV_1~.
\label{wtreedp2}
\ee

Clearly, there is a number of different cases we can consider. Besides the
generic case, where we have 4 gauge groups at the end of the cascade, 
we have particular cases with less gauge groups if we have 0 or 1 
branes of one or both kinds.

It is easy to see that adding only deformation branes ($K=0$), 
at low energies we basically get two decoupled $SU(P)$ SYM theories, 
together with baryonic branches which can be seen analyzing the next-to-last
step of the cascade, as in the conifold case \cite{KS}.

\begin{figure}[ht]
\begin{center}
{\includegraphics{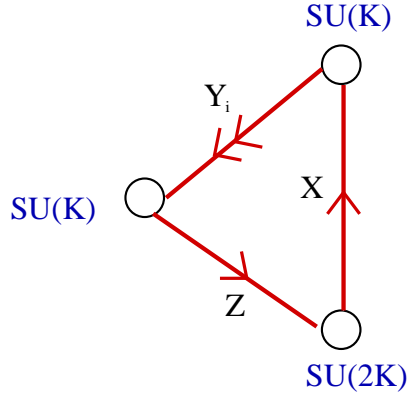}}
\caption{\small Quiver of the $dP_2$ theory for generic $K$ and $P=0$.}
\label{kldp2}
\end{center}
\end{figure}

On the other hand, in the presence of only SB branes ($P=0$), we end up
with a triangular quiver, as in Figure \ref{kldp2}. 

This can be shown to
have a runaway behaviour along the same lines as in the triangle
at the bottom of the $dP_1$ cascade discussed in the previous section.
The R-charges are listed in the table below.
\begin{table} [ht]
\label{Rcdp2}
\begin{center}
\begin{tabular}{c|c|c|c|c|}
 & $X$ & $Z$ & $Y_1$ & $Y_2$ \\
\hline
$U(1)_R$ & -1  & -1 & 4 & 0
\end{tabular}
\end{center}
\end{table}

The difference here is that in this case the R-charges of the baryonic invariants 
$(Y_1)^k(Y_2)^{M-k}$ which parameterize the classical flat directions,
are $R_k=4k$, and hence never equal to 2, no matter which number of fractional branes 
we have. Therefore, terms which could lift the flat directions would also break the 
R-symmetry and possibly lead to SUSY vacua, not to DSB. This can be checked by direct 
computations in a few low rank cases.

In a similar fashion, also the cases with generic $P$ and $K$, and with
$K$ generic\footnote{If $K=1$ and $P=1$, we have two $SU(2)$ gauge factors, 
each with 4 doublet matter fields and additional singlet fields. (Un)luckily,
the superpotential is not of IYIT type \cite{iy,it} and the usual R-charge
considerations show that this case should not lead to DSB. 
} and $P=1$ can be shown not to allow the lifting 
of classical flat directions while preserving the R-symmetry.

We thus turn to the only remaining case, that is the one where we have a (possibly
large) number $P$ of deformation branes, to which we add a single $K=1$ SB brane. 
As we have seen before, we found DSB in the $dP_1$ case by deforming
the theory for a single fractional brane to the well-known 3-2 model.
Here we will also be in a setup where a known model of DSB is recovered.
Let us then go through a more detailed analysis of this model.

The simplest generalization of the 3-2 model of \cite{ads} is to replace the
$SU(3)$ group by $SU(N)$ \cite{it2}, 
with $N$ odd to prevent the $SU(2)$ global anomaly.
Nothing really differs from the 3-2 model, except for the fact that the
gauge dynamics remains strongly coupled in the IR because an unbroken
$SU(N-2)$ gauge group survives. A slight modification of this model was presented in \cite{pst}. The
$SU(2)$ group has $N$ extra doublets instead of only one. Still, it
turns out that DSB is possible again only if $N$ is odd. The reason
is that one is led to add mass terms for all but one of the extra
doublets, but these mass terms are really ``baryonic'' terms of
$SU(2)$ and hence the mass matrix can have a single zero eigenvalue
only if it has odd dimensions.\footnote{In \cite{pst} the
$SU(N)-SU(N-2)$ model is also analyzed, which is obtained by
performing a Seiberg duality on the $SU(2)$. Of course the
conclusions about the IR behavior must be the same, but the
description is in a sense under better control (i.e. weakly coupled).}

\begin{figure}[ht]
\begin{center}
{\includegraphics{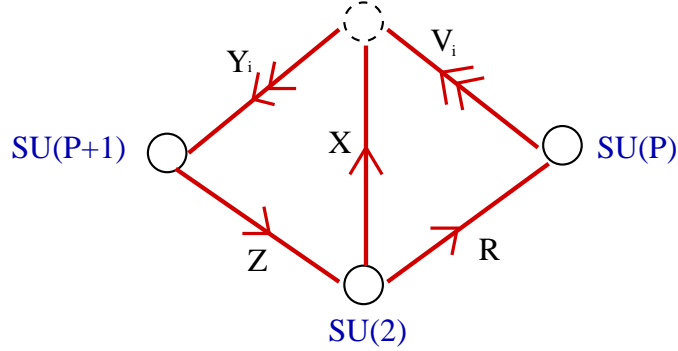}}
\caption{\small Quiver of the $dP_2$ theory for generic $P$ and $K=1$. The dashed
node represents an ``SU(1)'' factor, which plays no r\^ole in the IR.}
\label{ldp2}
\end{center}
\end{figure}

We now come to the analysis of the model at the base of a cascade
over $dP_2$ with $P$ deformation fractional branes and $K=1$ SB
fractional branes. The gauge group consists of 3 nodes:
$SU(P+1)\times SU(2) \times SU(P)$, and the matter content is given
in the quiver in figure \ref{ldp2} with the charge assignment in the
table below (note that we have used that
$\mathbf{2}=\mathbf{\overline 2}$, and we have included R-charges
that will be discussed later on).
\begin{table} [ht]
\begin{center}
\begin{tabular}{c|c|c|c|c|}
 & $SU(P+1)$  & $SU(2)$ & $SU(P)$ & $U(1)_R$\\
\hline &&&&\\
$Y_1$ & $\overline{\mathbf{P+1}}$  & $1$ & $1$ & $4-2P$\\
$Y_2$ & $\overline{\mathbf{P+1}}$  & $1$ & $1$ & $2$\\
$Z$ & $\mathbf{P+1}$  & $\mathbf{2}$ & $1$ & $-2$\\
$X$ & $1$  & $\mathbf{2}$ & $1$ & $2P$\\
$R$ & $1$  & $\mathbf{2}$ & $\overline{\mathbf{P}}$ & $2$\\
$V_1$ & $1$  & $1$ & $\mathbf{P}$ & $0$\\
$V_2$ & $1$  & $1$ & $\mathbf{P}$ & $-2P$
\end{tabular}
\end{center}
\end{table}

The expected dimension of the moduli space (in the absence of a tree
level superpotential) is given as follows. We have $4(P+1)+4P+2=8P+6$
matter fields, and by giving VEVs to all of them we are higgsing
$(P+1)^2-(P-1)^2+3+P^2-(P-2)^2=8P-1$ gauge fields, leaving a total of
$7$ classical flat directions. The gauge invariants we can build are 
$$
a_i=Y^a_iZ^\alpha_a X^\beta\epsilon_{\alpha\beta}~, \qquad
b_{ij}=Y^a_iZ^\alpha_a R^\beta_m V^m_j \epsilon_{\alpha\beta}~, \qquad
c=Z^\alpha_a Z^\beta_b Y^a_i Y^b_j \epsilon_{\alpha\beta}\epsilon^{ij},
$$
\be
d_i=X^\alpha R^\beta_m V^m_i \epsilon_{\alpha\beta}~, \qquad
e=R^\alpha_m R^\beta_n V^m_i V^n_j\epsilon_{\alpha\beta}\epsilon^{ij}~.
\ee
Note that the invariants $d_i$ are cubic only because the middle
node is $SU(2)$ and not of higher rank (in general they are baryons involving $3K$ 
fields). The above invariants are 10, but we also have the 3 constraints, 
\be
cd_i=a_j b_{ki}\epsilon^{jk}, \qquad
ce=b_{ij}b_{kl}\epsilon^{ik}\epsilon^{jl}.
\ee

The theory at the base of the cascade comes with the classical tree level
superpotential
\be
W_\mathrm{tree}=-a_1+b_{21} = -Y_1ZX+Y_2ZRV_1~.
\label{wtree1}
\ee
We can easily work out the F-flatness conditions, and find that the invariants
$a_i, b_{ij}$ and $c$ are set to zero while the $d_i$ and $e$ are left
undetermined.

Thus the latter are the light fields at low energies, and at a
generic point of the moduli space the fields $X, R$ and $V_i$ get VEVs.
This in particular implies, through (\ref{wtree1}), that the fields
$Y_i$ and $Z$ get masses, with a mass matrix which is such that
$\det m\propto d_1$. At low energies, the gauge group is broken
to $SU(P+1)\times SU(P-2)$ and the scale matching goes as follows
\be
\Lambda_{P+1,\mathrm{low}}^{3P+3}=h d_1 \Lambda_{P+1}^{3P+1}~,\quad
\Lambda_{P-2,\mathrm{low}}^{3P-6}=\frac{\Lambda_{P}^{3P-2}}{e}~,
\ee
where $h$ is the coupling of the quartic term in $W_\mathrm{tree}$,
reintroduced here for dimensional reasons.
The effective superpotential is thus
\be
W_\mathrm{eff}= (P+1) \left(h d_1 \Lambda_{P+1}^{3P+1}\right)^{\frac{1}{P+1}}
+(P-2)\left(\frac{\Lambda_{P}^{3P-2}}{e}\right)^{\frac{1}{P-2}}.
\label{weff}
\ee
It is clearly runaway, both in $d_1$ and in $e$.
The K\"ahler potential for the light modes can be computed, and it is given
by exactly the same expression as the one given in \cite{ads}
for the 3 moduli of the 3-2 model. Note that the model at hand
is similar to two $SU(N)-SU(2)$ models sharing the same $SU(2)$ node.
However, it turns out $P$ is not constrained to be even or odd.

As in the 3-2 model, it seems now plausible that adding a tree level term
including $d_2$ should lead to DSB. Though it would be nice to check this
using (\ref{weff}) and the K\"ahler potential, it is easier to analyze
the theory using the fundamental fields and their F-flatness conditions. 
We thus consider
\be
W_\mathrm{def}=W_\mathrm{tree}+d_2=-Y_1ZX+Y_2ZRV_1+XRV_2~.
\ee
Now the
classical F-conditions set all the invariants to zero, so that we
are in a situation where there is no classical moduli space. A first
indication that we should have DSB is that there is a non-anomalous
R-symmetry (whose charges are given in the table) which is preserved
by the above superpotential. Since (\ref{weff}) will push some
invariants to have non zero VEVs, the R-symmetry will be
spontaneously broken, thus SUSY must be broken, too.

We now write the effective superpotential including the non-perturbatively
generated ADS superpotential for both $SU(P+1)$ and $SU(P)$ gauge groups
\be
W_\mathrm{eff}=W_\mathrm{def}+(P-1)\left(\frac{\Lambda_{P+1}^{3P+1}}
{Z^2Y_1Y_2}\right)^{\frac{1}{P-1}}
+(P-2)\left(\frac{\Lambda_{P}^{3P-2}}{R^2V_1V_2}\right)^{\frac{1}{P-2}}.
\ee
The F-flatness conditions with respect to $X, V_i$ and $Y_i$ impose
that $a_2=b_{11}=b_{12}=b_{22}=c=d_1=e=0$ and that
$a_1=b_{21}=d_2\propto c^{-1/(P-1)} \rightarrow \infty$.
We now impose the F-flatness with respect to $R^\alpha_m$ and get
\be
(Y_2Z)^\alpha V_1^m+X^\alpha V_2^m=\left(\frac{\Lambda_{P}^{3P-2}}{e}
\right)^{\frac{1}{P-2}}\frac{1}{e}\left[(RV_1)^\alpha  V_2^m
-(RV_2)^\alpha  V_1^m \right]~.
\ee
Multiplying by $(X^\gamma\epsilon_{\alpha\gamma})
(R^\beta_nV^n_2R^\delta_m\epsilon_{\beta\delta})$, we get
\be
a_2 e =\left(\frac{\Lambda_{P}^{3P-2}}{e}\right)^{\frac{1}{P-2}} d_2~,
\ee
which is an inconsistency since the l.h.s. is supposed to vanish while
the r.h.s. should blow up. Hence, the F-flatness conditions have no solution,
even for infinite VEVs. We conclude that this model displays true DSB and a stable
non-supersymmetric vacuum.

Again, the coupling that we had to introduce to stabilize the runaway behaviour
uses the fact that, due to the $SU(2)$ group, 
we can turn around the arrow for the field $X$, and the coupling is baryonic
in this sense.

Let us stress the basic difference between the present case and the $dP_1$ case.
Here we have $P$ deformation branes which provide the non-perturbative
correction to the superpotential. This translates to the smooth deformation
of the gravity dual, which should be (for $K=0$) a KS-like geometry.
Then comes the addition of the single SB fractional brane, which could in
principle be treated in the probe approximation, also because all the
effects it has on the dynamics are through tree level couplings in the
superpotential and the appearance of an $SU(2)$ gauge group which is IR-free
for sufficiently large $P$. Hence, we expect that it could be possible
to see both the runaway behaviour, with the original $W_\mathrm{tree}$, and
a stable non-supersymmetric configuration, with $W_\mathrm{def}$, on the world-volume
of the probe fractional brane in the KS-like background produced by the 
$P\gg1$ deformation branes through geometric transition.

\subsection{Comments on the $X^{p,q}$ Family}

The second del Pezzo surface can be obtained as a blow-up of the first del Pezzo. The field theory dual 
of this geometric operation was shown in \cite{Feng:2002fv} to correspond to an unhiggsing procedure, i.e., 
the superconformal gauge theory dual to $dP_2$ can be obtained by unhiggsing from that of $dP_1$. Based on this, 
in \cite{Hanany:2005hq} a new family of toric singularities, dubbed $X^{p,q}$, was constructed and 
the corresponding dual gauge theories derived. A SE manifold $X^{p,q}$ 
is related to $Y^{p,q}$ 
right in the same way as $dP_2$ is related to $dP_1$. And a similar relation holds for the corresponding 
dual gauge theories. For instance, the gauge theory dual to $X^{p,q}$ has one gauge group more than the 
corresponding $Y^{p,q}$ theory.\footnote{To find the explicit metrics of the $X^{p,q}$ is still an open problem.} 

The $X^{p,q}$ theories admit two kinds of fractional branes, similarly to $dP_2$, one of them being of the SB 
type and leading, in the IR, to a runaway behaviour. Once again, one could ask if the runaway can be stabilized  
by some suitable deformation of the superpotential. Again, we suggest the answer to be negative. Explicit 
computations for some simple cases, and general considerations similar to those holding for $Y^{p,q}$, 
suggest that as long as $p$ and $q$ increase,  the number of classical 
flat directions increases faster then the number of possible ``stabilizing deformations'' to be added to the 
superpotential \cite{ABC}.

\section{Stable Vacua at the Bottom of the $\mathbf{dP_3}$ Cascade}

One might ask if the pattern of DSB we have described for $dP_2$ goes through as we go on 
with higher del Pezzo surfaces, which are in fact all related by subsequent blow-ups. The analysis for the 
next del Pezzo, $dP_3$, is straightforward. This 
theory and the corresponding cascade was already studied in the literature. We refer to 
\cite{Franco:2004jz,FHU,FHSU} for more details. 

\begin{figure}[ht]
\begin{center}
{\includegraphics{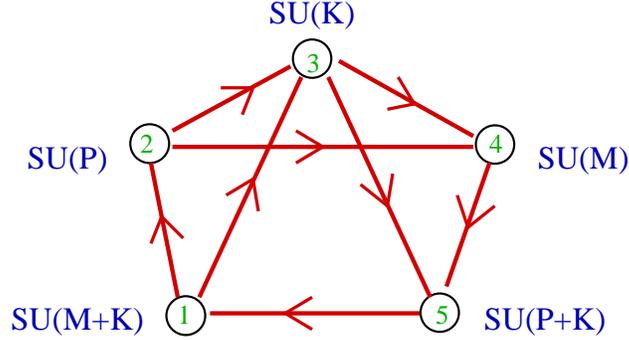}}
\caption{\small Quiver of the $dP_3$ theory for generic $M,P$ and $K$.}
\label{ldp3g}
\end{center}
\end{figure}

The $dP_3$ theory admits three different independent fractional branes, all of the deformation type. Two of 
them, similarly to the deformation brane of the $dP_2$ case, lead in the IR to two decoupled nodes. The third 
one leads to a triangle quiver with the corresponding cubic gauge invariant being present in the tree level 
superpotential (resulting, however, in the same confining dynamics). The theory at the bottom of the cascade, 
for generic values of the three fractional branes, 
$M,P$ and $K$, respectively, is clearly more complicated. It is depicted in Figure \ref{ldp3g} and has a tree 
level superpotential 
\be
\label{w3}
W = X_{13}X_{35}X_{51} - X_{12}X_{24}X_{45}X_{51}~.
\ee
As shown in \cite{FHSU} these deformation branes are not mutually supersymmetric. More precisely, for generic $M$ 
and $P$ but $K=0$, the cascade does end in a supersymmetric confining theory, while, whenever $K$ and $M$ (and/or $P$) 
are simultaneously different from zero, then an ADS superpotential 
is generated (this should be clear from the quiver in Figure \ref{ldp3g}). Take, for instance, $P=0$ and generic 
$M$ and $K$. In this case the theory reduces to four gauge groups with group 1 having $N_F<N_C$ and a superpotential 
with only the cubic coupling
\be
W = X_{13}X_{35}X_{51}~.
\ee

\begin{figure}[ht]
\begin{center}
{\includegraphics{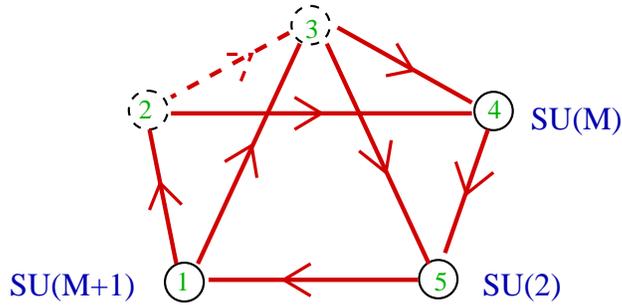}}
\caption{\small Quiver of the $dP_3$ theory for generic $M$ and $P=K=1$.
The dashed nodes represent ``SU(1)'' factors while the dashed line a bi-fundamental field which completely
decouples in the IR and plays no r\^ole. This theory is essentially the same as the one for $dP_2$, see 
Figure \ref{ldp2}.}
\label{ldp3}
\end{center}
\end{figure}

The dynamically generated ADS superpotential leads to runaway. Indeed, for generic values of $M,P$ and $K$ 
there are plenty of classical flat directions, which are very difficult to lift altogether. A general analysis based 
on possible R-charge preserving superpotential couplings singles out only one possibility which allows for a 
deformation which lifts all classical flat directions. This is for generic $M$ and $P=K=1$ (or equivalently generic $P$ 
and $M=K=1$) see Figure \ref{ldp3}. 

Note that since the ``$SU(1)$'' gauge factors actually do not exist, 
the IR theory is nothing but the one at work 
for the $dP_2$ case, Figure \ref{ldp2}. Hence, the analysis is exactly the same as the one in the 
previous section, as well as the 
conclusions. This seems to indicate that some sort of unique picture emerges from (toric) del 
Pezzo's. The analysis in the next section will support this claim.

Similarly to the case of the first and the second del Pezzo surfaces, one could argue the existence of a full family 
of new (toric) SE manifolds, call them $Z^{p,q}$, with three different kinds of fractional branes, whose dual gauge theories 
should be obtained by unhiggsing from those dual to $X^{p,q}$. In fact, it should be possible to construct such 
new dual pairs, along the same lines of \cite{Hanany:2005hq} for $X^{p,q}$, using toric geometry. We expect that similar 
conclusions as those for the $Y$ and $X$ families hold for the $Z$ family, i.e. no possible stabilization of the otherwise 
runaway behaviour whenever $p>2\,,\,q>1$.

\section{Higher (Pseudo) del Pezzo's}

\begin{figure}[ht]
\begin{center}
{\includegraphics{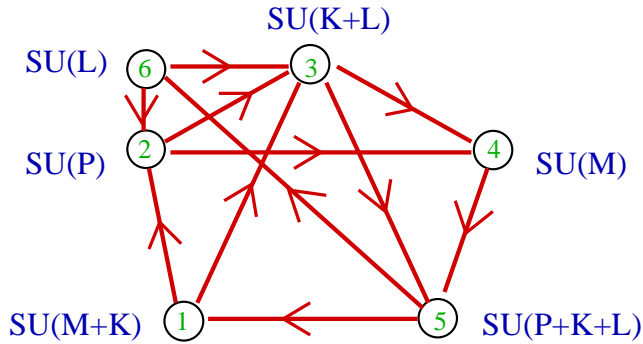}}
\caption{\small Quiver of the $PdP_4$ theory for generic $M,P,K$ and $L$.}
\label{lpdp4g}
\end{center}
\end{figure}

By further unhiggsing one can go up in the number of gauge groups and get other dual toric 
varieties by the geometric dual blow-up procedure \cite{Feng:2002fv}. These were dubbed Pseudo del Pezzo, 
$PdP_k$, not to be confused with actual del Pezzo's which, for $k>3$, are not toric. We consider here, as an example, 
the $PdP_4$ theory. This model admits four different kinds of fractional branes \cite{FHSU}. Three are deformation branes and 
pair exactly those of $dP_3$. The last is a ${\cal N}=2$ brane (it leads to a triangle quiver in the IR with the 
corresponding cubic gauge invariant {\it not} being present in the tree level superpotential). For generic numbers $M,P,K$ 
and $L$ of these fractional branes, respectively, the quiver at the bottom of the cascade is as in Figure \ref{lpdp4g},

with a superpotential
\be
\label{w4}
W = X_{12}X_{24}X_{45}X_{51} -  X_{63}X_{34}X_{45} X_{56} + X_{35}X_{56}X_{62}X_{23} - X_{13}X_{35}X_{51}~.
\ee

Suppose to consider no ${\cal N}=2$ branes first, i.e. $L=0$. It is easy to see that the quiver in Figure \ref{lpdp4g} 
reduces exactly to that of $dP_3$, Figure \ref{ldp3g} (just drop node 6 and the corresponding matter fields) and
the superpotential (\ref{w4}) reduces to the superpotential (\ref{w3}). Therefore the analysis 
in this case is the same as before: the {\it only} possibility which allows for a stabilization of the runaway behaviour 
is taking $M$ generic (and possibly large) and $P=K=1$, as we did in the case of $dP_3$. 
Once again, as far as DSB is concerned, the physics resembles that of $dP_2$: only in one specific 
case DSB is possible, providing a model resembling very much the $SU(N)-SU(2)$ model. 

We are left to consider the case where $L\not =0$. The analysis gets clearly more involved and we 
limit here to few basic observations. Depending on the relative value of $M,P,K$ and $L$ there 
are indeed ADS superpotential terms being generated and possible DSB. However, the presence of $L$ makes 
the appearance of an extra gauge invariant, $ X_{35}X_{56}X_{63}$, 
which is {\it not} present in the superpotential. 
Hence, for $L \not =0$, on top of the flat baryonic directions there will also be additional
mesonic flat directions, which makes it even more unlikely to find DSB by
adding simple terms to the superpotential. 

\section{Discussion}

In this paper we have presented a few instances of cascading quiver
gauge theories where, by deforming the tree level superpotential at the
bottom of the cascade by simple ``baryonic'' terms, we could find genuine
dynamical supersymmetry breaking in a stable vacuum. The pattern
which seems to emerge is that we can find one such instance in each
set of quiver gauge theories dual to branes on the complex cone over
del Pezzo surfaces. Except for the case of $dP_1$, it is
possible to have a large number of deformation branes and thus treat
the addition of the other branes which lead to SUSY breaking in the probe
approximation. 

We have also argued that we do not expect the same mechanism to work in
the infinite families of quiver gauge theories which are generalizations
of the ones corresponding to
del Pezzo surfaces, as the $Y^{p,q}$ and $X^{p,q}$.
In this line of reasoning, it would be interesting to use more systematic
methods to address these questions in general \cite{ABC}, 
taking also into account the relationship among these different theories
by higgsing and unhiggsing \cite{Feng:2002fv}. Though we expect
that the outcome of this more general study will be more of the kind of
a ``no-go'' theorem except for the known cases, there is still the possibility
that other kinds of models of DSB would appear to be relevant.
With respect to this possibility, we think that it is nevertheless
likely that it will always be necessary to deform in some way the
theory in order to get DSB. Indeed, though a definite theorem still does not
exist, there is convincing evidence \cite{Brini:2006ej} that the generic
behaviour when fractional branes of any kind are added to a conformal
quiver gauge theory is that we end up at the end of the cascade
with either SUSY vacua, or a runaway behaviour. This is essentially
because ``baryonic'' invariants are always present at the bottom
of the cascade, and they are always left as classical flat directions
by the tree level superpotential inherited from the conformal quiver.

It would also be interesting to extend these ideas to brane systems
probing non-toric geometries, such as for instance the $dP_{n>3}$ surfaces,
which have remained elusive to this point.

We should pay a closer attention to the terms that we add to the tree level
superpotential in order to have DSB, terms that we generically dub 
``baryonic''. These terms can be written because at the bottom of the
cascade an $SU(2)$ gauge factor arises and arrows can be reversed.
An interesting question is to see how these terms would look like if we
go up the cascade, or in other words what their origin can be in the
theory far from the IR, where a large (effective) number of
regular branes is present. Though performing a detailed analysis is beyond the scope
of the present work, we expect these operators to correspond to 
(combinations of) baryonic-like operators, of which there
are plenty at a generic step in the cascade. Under Seiberg dualities,
baryonic invariants and even baryonic deformations change simply by
a redefinition of their constituent fields \cite{aharony1,aharony2}.
Hence we expect the cascade, and its self-similarity, 
not to be affected by these operators, which are essentially IR irrelevant
at least until the last step of the cascade.

In the present paper we have adopted a pure gauge theory approach to the
problem. It would be extremely interesting to study the gravity/string
counterpart of the examples we consider.\footnote{See \cite{deBoer:1998by,Maldacena:2001pb,Diaconescu:2005pc} 
for the study of other models of DSB from the string/gravity/brane perspective.} 
It is interesting to note that the probe approximation seems to be imposed on us by 
the problem itself. It would be nice to understand from a string theory point of view
what is special for a single fractional brane, that cannot be generalized
to a bunch $M>1$ of the same kind of fractional branes. This is of course
in a way related to the enlarged flavour group of $SU(2)$ SQCD with respect
to $SU(N_c>2)$, a fact that to our knowledge has not been really dealt with
in the literature on gauge theories from brane setups.

The next issue would be to understand how to generate the extra terms
in the superpotential in string theory. Baryonic invariants of a CFT 
in the AdS/CFT correspondence are usually associated to D3 branes 
wrapped on 3-cycles \cite{gubser,intri,herzog1,herzog2}. From this
point of view it would seem rather difficult to write deformations
of the theory by such operators. However, from the probe brane point of
view, the operators are more innocent-looking, quadratic or cubic as
they are. Hence we have at least the hope that, as in the usual
treatment of flavours \cite{Karch}, deformations involving matter fields
can be implemented on the world-volume theory of the probe brane.

We would like to end this discussion with more ``phenomenological'' issues.
It would be very interesting to use the gauge/string correspondence
to study the spectrum of the theories which display DSB. Again, because
of the probe approximation, the study in the string/gravity dual
should be along the lines of the one performed for theories with flavours.

We should be able to observe a spectrum characterized by the scale of
DSB, and of course the goldstino, i.e. the Goldstone fermion of broken SUSY.
Note that we expect the goldstino not to correspond to a bulk 
(closed string) mode, as the one discussed in \cite{Argurio:2006my}, but
rather to a mode in the probe brane world-volume theory.
This does not prevent the existence of other massless modes in the (supersymmetric) bulk,
as in the conifold theory (see \cite{ghk} for the bosonic modes and
\cite{Argurio:2006my} for their fermionic partners), associated to quantum flat
directions in the gauge theory and possibly to Goldstone bosons of broken
symmetries. The latter remain of course exactly massless even if SUSY is
broken, however we expect in the probe approximation the other modes
to remain massless too. In a full back-reacted geometry they would presumably 
be lifted.

\vskip 15pt
\centerline{\bf Acknowledgements}
\vskip 10pt
\noindent
We thank Francesco Benini, Sergio Benvenuti, Francesco Bigazzi, Giulio Bonelli, Andrea Brini, Aldo Cotrone, 
Gabriele Ferretti, Davide Forcella, Dario Martelli, Angel Uranga and Alberto Zaffaroni for very useful discussions 
and/or email correspondence. We are particularly grateful to Gabriele Ferretti and Alberto Zaffaroni for reading a 
preliminary version of the manuscript, and to Davide Forcella for clarifying discussions about toric geometry 
and dimer technology. This work is partially supported by the European Commission FP6
Programme MRTN-CT-2004-005104, in which R.A and C.C. are associated to
V.U. Brussel. R.A. is a Research Associate of the Fonds National de la Recherche
Scientifique (Belgium). The research of R.A. and C.C. is partially supported by IISN - Belgium 
(convention 4.4505.86) and by the ``Interuniversity Attraction Poles Programme --Belgian Science Policy''. 
M.B. and S.C. are partially supported by Italian MIUR under contract PRIN-2005023102. M.B. is also supported by a 
MIUR fellowship within the program ``Rientro dei Cervelli''.


\end{document}